\documentclass[useAMS,usenatbib]{mn2e}
\usepackage{mn2e-breakabs}
\usepackage{graphicx}
\usepackage{subfigure}
\usepackage{times}
\usepackage{caption}
\usepackage{rotating}
\usepackage{rotfloat}
\usepackage{multicol}
\usepackage{array}
\usepackage{booktabs}
\usepackage{color}
\usepackage{afterpage}
\usepackage[multidot]{grffile}
\usepackage{bm}
\usepackage{float}
\usepackage{hyperref}
\providecommand{\adsurl}[1]{\href{#1}{ADS}}

\newcommand{\hompc}{\,h\,{\rm Mpc}^{-1}}
\newcommand{\mpcoh}{\,h^{-1}\,{\rm Mpc}}
\newcommand{\Hunit}{\,{\rm km}\,{\rm s}^{-1}\,{\rm Mpc}^{-1}}

\def\la{\mathrel{\mathpalette\fun <}}
\def\ga{\mathrel{\mathpalette\fun >}}
\def\fun#1#2{\lower3.6pt\vbox{\baselineskip0pt\lineskip.9pt
        \ialign{$\mathsurround=0pt#1\hfill##\hfil$\crcr#2\crcr\sim\crcr}}}

\newcommand{\be}{\begin{equation}}
\newcommand{\ee}{\end{equation}}
\newcommand{\ba}{\begin{eqnarray}}
\newcommand{\ea}{\end{eqnarray}}
\newcommand{\simgt}{\,\hbox{\lower0.6ex\hbox{$\sim$}\llap{\raise0.6ex\hbox{$>$}}}\,}
\newcommand{\simlt}{\,\hbox{\lower0.6ex\hbox{$\sim$}\llap{\raise0.6ex\hbox{$<$}}}\,}

\begin{document}

\title[EZmock]
{
EZmocks: extending the Zel'dovich approximation to generate mock galaxy catalogues with accurate clustering statistics
}

\author[Chuang et al.]{
  \parbox{\textwidth}{
 Chia-Hsun Chuang$^1$\thanks{E-mail: chia-hsun.chuang@uam.es, MultiDark Fellow},
 Francisco-Shu Kitaura$^2$\thanks{E-mail: kitaura@aip.de, Karl-Schwarzschild-Fellow},
 Francisco Prada$^{1,3,4}$,
 Cheng Zhao$^5$,
 Gustavo Yepes$^{6}$,
}
  \vspace*{4pt} \\
$^1$ Instituto de F\'{\i}sica Te\'orica, (UAM/CSIC), Universidad Aut\'onoma de Madrid,  Cantoblanco, E-28049 Madrid, Spain \\
$^2$ Leibniz-Institut f\"ur Astrophysik Potsdam (AIP), An der Sternwarte 16, D-14482 Potsdam, Germany\\
$^3$ Campus of International Excellence UAM+CSIC, Cantoblanco, E-28049 Madrid, Spain \\
$^4$ Instituto de Astrof\'{\i}sica de Andaluc\'{\i}a (CSIC), Glorieta de la Astronom\'{\i}a, E-18080 Granada, Spain \\
$^5$ Tsinghua Center for Astrophysics, Department of Physics, Tsinghua University, Haidian District, Beijing 100084, P. R. China\\
$^6$Departamento de F{\'i}sica Te{\'o}rica,  Universidad Aut{\'o}noma de Madrid, Cantoblanco, 28049, Madrid, Spain\\
}

\date{\today} 

\maketitle

\begin{abstract}
We present a new methodology to generate mock halo or galaxy catalogues, which have accurate clustering properties, nearly indistinguishable from full $N$-body solutions, in terms of the one-point, two-point, and three-point statistics. In particular, the agreement is remarkable, within $1\%$ up to $k=0.55$ $\hompc$ and down to  $r=10$ $\mpcoh$, for the power spectrum and two-point correlation function respectively, while the bispectrum agrees in general within $20\%$ for different scales and shapes. Our approach is based on the Zel'dovich approximation, however, effectively including with the simple prescriptions the missing physical ingredients, and stochastic scale-dependent, non-local and nonlinear biasing contributions. The computing time and memory required to produce one mock is similar to that using the log-normal model. With high accuracy and efficiency, the effective Zel'dovich approximation mocks (EZmocks) provide a reliable and practical method to produce massive mock galaxy catalogues for the analysis of large-scale structure measurements.
\end{abstract}

\begin{keywords}
 cosmology: observations - distance scale - large-scale structure of
  Universe
\end{keywords}

\section{Introduction} \label{sec:intro}

The scope of galaxy redshift 
surveys has dramatically increased in the last decade. The 2dF Galaxy Redshift Survey\footnote{http://www2.aao.gov.au/2dfgrs/} (2dFGRS) 
obtained 221,414 galaxy redshifts at $z<0.3$ \citep{Colless:2001gk,Colless:2003wz}, 
and the Sloan Digital Sky Survey\footnote{http://www.sdss.org} (SDSS, \citealt{York:2000gk}) collected 
930,000 galaxy spectra in the Seventh Data Release (DR7) at $z<0.5$ \citep{Abazajian:2008wr}.
WiggleZ\footnote{ http://wigglez.swin.edu.au/site/} collected spectra of 240,000 emission-line galaxies at $0.5<z<1$ over 
1,000 square degrees \citep{Drinkwater:2009sd, Parkinson:2012vd}, and the Baryon Oscillation Spectroscopic Survey\footnote{https://www.sdss3.org/surveys/boss.php} (BOSS, \citealt{Dawson:2012va}) of the SDSS-III project\citep{Eisenstein:2011sa} is surveying 1.5 million luminous red galaxies (LRGs) at $0.1<z<0.7$ over 10,000 square degrees.
There are new upcoming ground-based and space experiments, such as
4MOST\footnote{http://www.4most.eu/} (4-metre Multi-Object Spectroscopic Telescope, \citealt{deJong:2012nj}), DES\footnote{http://www.darkenergysurvey.org} (Dark Energy Survey), DESI\footnote{http://desi.lbl.gov/} (Dark Energy Spectroscopic Instrument,\citealt{Schlegel:2011zz,Levi:2013gra}), eBOSS\footnote{http://www.sdss.org/sdss-surveys/eboss/} (Extended Baryon Oscillation Spectroscopic Survey), 
HETDEX\footnote{http://hetdex.org} (Hobby-Eberly Telescope Dark Energy Experiment, \citealt{Hill:2008mv}), 
J-PAS\footnote{http://j-pas.org} (Javalambre Physics of accelerating universe Astrophysical Survey, \citealt{Benitez:2014ibt}), 
LSST\footnote{http://www.lsst.org/lsst/} (Large Synoptic Survey Telescope, \citealt{Abell:2009aa}), 
Euclid\footnote{http://www.euclid-ec.org } \citep{Laureijs:2011gra}, 
and WFIRST\footnote{http://wfirst.gsfc.nasa.gov} (Wide-Field Infrared Survey Telescope, \citealt{Spergel:2013}).

Mock galaxy catalogues are essential for analysing the clustering signal drawn from these surveys. The most reliable technique for creating mock catalogues is $N$-body cosmological simulations, e.g., LasDamas\footnote{http://lss.phy.vanderbilt.edu/lasdamas/} (Large Suite of Dark Matter Simulations), which has been used to analyse the SDSS-II galaxy sample (e.g., \citealt{Chuang:2010dv, Samushia:2011cs}). 
However, the total run-time and memory required to generate a large suit of simulations make this effort prohibitive in most of the cases, hence their use for ongoing and future surveys is impractical. 
Recent techniques permit to speed up $N$-body codes (see, e.g., COLA\footnote{COLA (COmoving Lagrangian Acceleration simulation)}: \citealt{Tassev:2013pn}), however, the memory requirements are still large with these methods.

Alternatively, log-normal mock catalogues \citep{Coles:1991if} have been widely used (e.g., \citealt{Cole:2005sx,Percival:2009xn,Reid:2009xm,Blake:2011wn,Beutler:2011hx}). 
The log-normal model produces by construction mock catalogues with precise two-point statistics, transforming Gaussian density fields with a given variance. Yet, its co-moving description neglects the relative displacement of matter through cosmic evolution, and does not capture the spatial pattern of the cosmic web. Hence, it leads to large deviations in the higher-order statistics of the large-scale structure.

More realistic mock catalogues can be produced describing the gravity induced motion of matter with perturbation theory. Two codes have pioneered the field based on this technique, namely PTHalos \citep{Scoccimarro:2001cj} and PINOCCHIO\footnote{PINOCCHIO (PINpointing Orbit-Crossing Collapsed Hierarchical Objects)} \citep{Monaco:2001jg,Monaco:2013qta}. More recently, PTHalos has been applied for the analysis of the BOSS galaxy survey \citep{Manera:2012sc,Manera:2014cpa}. The accuracy of these methods is limited by the approximate solutions of perturbation theory, unable to model the nonlinear regime towards small scales, and hence, to accurately resolve the formation of haloes hosting the galaxies. 
To overcome this problem, recent approaches, relying on a statistical description of the halo distribution have been introduced. These methods require only the large scale density field and therefore can rely on approximate gravity solvers, either based on improved perturbation theory algorithms (see PATCHY\footnote{PATCHY (PerturbAtion Theory Catalog generator of Halo and galaxY distributions)}: \citealt{Kitaura:2013cwa,Kitaura:2014mja}), or on approximate particle mesh solutions (see QPM\footnote{QPM (quick particle mesh)}: \citealt{White:2013psd}).  This density field is populated with haloes or galaxies in a posterior step. 
We want to further explore this approach, reducing the computational requirements to a minimum, albeit without losing  accuracy.

In this work, we develop a new methodology  to generate mock halo or galaxy catalogues, which have accurate clustering properties, nearly indistinguishable from full $N$-body solutions, in terms of the one-point, two-point, and three-point statistics. Our approach is based on the Zel'dovich approximation (ZA, \citealt{Zeldovich:1969sb}), however, effectively including stochastic scale-dependent, non-local and nonlinear biasing contributions. 

The computation of the effective Zel'dovich approximation mock catalogues (EZmocks, hereafter) demands a minimum run-time, i.e., three fast Fourier transform (FFT) to compute the displacement field in three directions and populate haloes nearby each grid point with some random assignment process, and a minimum memory requirement, i.e., a few arrays of the same size as the grid used by FFT to store the information of the displacement field and the number of generated haloes. 
The ongoing and upcoming large galaxy surveys will demand precise mock catalogues scanning huge parameter space. Simple and efficient, but accurate methods will be required to cover those needs.
While this year celebrates the $100^{th}$ anniversary of the birth of Yakov Zel'dovich, we find that the Zel'dovich approximation can still provide one of the most powerful tools for analysing the large-scale structure survey data.

This paper is organised as follows. In Section \ref{sec:sim}, we describe the BigMultiDark $N$-body simulations and the computational facilities used for this study. In Section \ref{sec:method}, we provide the details of the 
methodology to construct the  EZmocks. 
In Section \ref{sec:results}, we show its performance by comparing with $N$-body simulations.
We summarise and conclude in Section \ref{sec:conclusion}.

\section{Reference simulations and computational requirements}
\label{sec:sim}
To test our method, we use a reference halo catalogue at redshift $z=0.5618$ extracted from one of the  BigMultiDark (BigMD) simulations\footnote{http://www.multidark.org/MultiDark/} (Klypin et al. in prep.), which was performed using \textsc{gadget-2} \cite{Springel:2005mi} 
with $3840^3$ particles on a volume of $(2500$ $\mpcoh)^3$ assuming $\Lambda$CDM Planck cosmology with \{$\Omega_{\rm M}=0.307115, \Omega_{\rm b}=0.048206,\sigma_8=0.8288,n_s=0.96$\}, and a Hubble constant ($H_0=100\,h\Hunit$) given by  $h=0.6777$. 
Haloes were defined based on two different algorithms. One identifies density peaks including substructures using the Bound Density Maximum (BDM) halo finder \citep{Klypin:1997sk,Riebe:2011gp} 
and the other one groups up the particles using a friend-of-friend (FOF) halo finder.
In this work, we use the FOF catalogue as our reference. 
From the halo catalogue, we select a complete sample, selected by mass, with number density $3.5\times 10^{-4}$ $h^3\,{\rm Mpc}^{-3}$ which is similar to that of the BOSS galaxy sample at $z\sim0.5$. 
The MultiDark simulations have been used to interpret the clustering of the BOSS galaxy sample (e.g., see \citealt{Nuza:2012mw}). 

We are using the thin-nodes of the Curie supercomputer\footnote{http://www-hpc.cea.fr/en/complexe/tgcc-curie.htm} to generate EZmocks. It has 16 cores (2.7GHz) with 64GB of memory per node (we only need less than 32GB in this study) and 5040 nodes in total. We use shared memory multiprocessing (i.e., OpenMP) to speed up the computation. It takes less than 5 minutes to construct one EZmock with the grid size of $960^3$ using one node. In other words, it takes 10 minutes to generate 10,000 EZmocks using full power of the machine. This make EZmock really competitive as compared to other approximation methods mentioned above, which certainly demand more computational resources.

\section{Methodology} 
\label{sec:method}

In this section, we introduce the methodology to construct  EZmocks. The ZA yields a crude approximation of the dark matter density field on small scales, which cannot be directly used to compute the halo or galaxy catalogue. Therefore, the different components, which are not modelled by the ZA, have to be added in subsequent steps. The main assumption we test in this study is that the ZA is sufficiently accurate on large scales to account for the three dimensional shape of the cosmic web. We aim at including with simple prescriptions the missing physical ingredients, such as, tidal fields (included in second order LPT), and nonlinear, non-local stochastic biasing. The following required steps are recursively applied until convergence:

\begin{enumerate}
 \item[(I)] generation of the dark matter density field on a grid using the Zel'dovich approximation (ZA);
\item[(II)] mapping the probability distribution function (PDF) of haloes measured in BigMD to the ZA density field;
\item[(III)] adding scatter to the PDF mapping scheme;
\item[(IV)] fitting the amplitude of the power spectrum and bispectrum with a density threshold and saturation; 
\item[(V)] fitting the shape of the final power spectrum by modifying the tilt in the initial input power spectrum with a scale-dependent function;
\item[(VI)] fitting baryon acoustic oscillations (BAOs) by enhancing the amplitude of BAOs in the initial input power spectrum;
\item[(VII)] computing the velocity field within the ZA for each object.
\end{enumerate}

The steps listed above have a physical analogy. While the scatter in step III emulates stochastic biasing, the combination of PDF mapping (step II), thresholding (step IV) and tilting of the power spectrum (step V) mimics the effects of nonlinear, deterministic biasing. Non-local effects, such as tidal fields not included in linear LPT or other biasing contributions, are effectively included in both, the scatter relation and the tilting of the initial power spectrum. The missing power towards large $k$s (small scales) of perturbative approaches, is included in the modulation of the initial power spectrum when fitting the resulting halo population. In this way, multiple effects are accounted for in a single procedure.

We note that the approach presented in this work focuses on fitting the clustering statistics with direct and explicit prescriptions, circumventing the problem of finding the parameters to a particular biasing model, like it is done for instance in the PATCHY code. This makes the parameter finding in EZmocks much more efficient at the expense of giving up more explicit physical relations.

\subsection{Generation of the Zel'dovich density field}
\label{sec:ZA}
A particle located at Lagrangian position $\boldsymbol q$ will be mapped to its Eulerian position $\boldsymbol x$ at cosmic time $t$ by the displacement field $\boldsymbol\Psi(\boldsymbol q,t)$, i.e.,
\begin{equation}
\boldsymbol x(\boldsymbol q,t)=\boldsymbol q+\boldsymbol\Psi(\boldsymbol q,t).
\end{equation}
The first order Lagrangian perturbation theory solution to the equations of motions is given by the Zel'dovich approximation (for a review, see, e.g., \citealt{Bernardeau:2001qr}).
The displacement field in the ZA is given by
\begin{equation}
\boldsymbol\Psi(\boldsymbol q)=\int{\frac{d^3k}{(2\pi)^3}e^{i\boldsymbol k\cdot \boldsymbol q}\frac{i\boldsymbol k}{k^2}\hat\delta(\boldsymbol k)},
\end{equation}
where $\hat\delta(\boldsymbol k)$ is the fractional density perturbation in Fourier-space.
We construct the displacement field using the ZA to the same redshift as the reference catalog, i.e., $z = 0.5618$. The adopted grid size, $960^3$, is a factor of 64 smaller, as compared to the one used by the BigMD simulation. 
We have tested that smaller grid sizes than the one presented in this study significantly damp the BAO peak. We have also verified that we get the same results with larger grid sizes, however, at a higher computational cost. Therefore, we consider that the resolution chosen in this study is very close to be the optimal one.
We use the reduced white noise from the BigMD simulation to reduce the cosmic variance throughout this work. The density field is obtained using the cloud-in-cells particle assignment scheme (CIC, e.g., \citealt{Hockney1981}).

\subsection{PDF mapping scheme}
\label{sec:PDF}
In this step, we map the probability distribution function measured from the halo catalog extracted from the $N$-body simulation to that from the ZA density field based on a rank ordering procedure. We note that this step is done on the halo field rather than on the dark matter field, and is not sufficient to solve the damping of power in perturbative approaches alone (e.g., see \citealt{Leclercq:2013kza}).

Our rank ordering procedure is as follows:
\begin{enumerate}
  \item[(a)] computing the halo density field from BigMD with CIC;
  \item[(b)] generating an integer on each grid point using a Poisson random number generator based on the density computed from step (a);
  \item[(c)] assigning these integers to the grid of our ZA density field by rank ordering;
  \item[(d)] populating the mock catalogue following a CIC distribution. The number of haloes populated on each grid point is the one assigned from step (c). 
The CIC distribution assigns halos only in the eight neighbor cells of the grid point. Assuming a position (in one of the eight neighbor cells) has relative distance $\{\Delta x, \Delta y, \Delta z\}$ to the grid point. For each halo, the probability that this position gets to be assigned the halo is proportional to $(\mbox{cell-size}-\mid\Delta x\mid)(\mbox{cell-size}-\mid\Delta y\mid)(\mbox{cell-size}-\mid\Delta z\mid)$.
It can be implemented by
\begin{equation}
\Delta x = 
\left\{ \begin{array}{ll}
         (1-\sqrt{R})\times\mbox{cell-size} & \mbox{if $R \ge 0$};\\
        (-1+\sqrt{-R})\times\mbox{cell-size} & \mbox{if $R < 0$},
\end{array} \right.
\end{equation}
where $\Delta x$ is the distance from the halo to the grid point in $x$ direction, $R$ is a random number drawn between $-1$ and $1$, and cell-size is the separation of the neighboring grid points, i.e. $2500/960\mpcoh$ in this work. $\Delta y$ and $\Delta z$ are determined with different draws of $R$.
\end{enumerate}

\subsection{Adding scatter to the PDF mapping}
\label{sec:scatter}
This step models the uncertainty in the PDF relation and hereby also the stochasticity of the tracers.
While mapping the PDF with strict rank ordering, one finds that the amplitude of the power spectrum of the mock catalog is too high. Therefore, we introduce a scatter to the ZA density field by 
\begin{equation}
\rho_{s}(\boldsymbol r) = 
\left\{ \begin{array}{ll}
         \rho_{o}(\boldsymbol r)(1+G(\lambda)) & \mbox{if $G(\lambda) \ge 0$};\\
        \rho_{o}(\boldsymbol r)\exp(G(\lambda)) & \mbox{if $G(\lambda) < 0$},
\end{array} \right.
\end{equation}
where $\rho_s(\boldsymbol r)$ and $\rho_o(\boldsymbol r)$ are the ZA density field after and before the scattering respectively. $G(\lambda)$ is a random number drawn from the Gaussian distribution with width $\lambda$. The exponential function is used to avoid the negative density. 
To simplify the problem, we fix $\lambda=10$. The choice of $\lambda$ will affect the values of the other parameters in our models but does not affect the performance of the mock catalogue.
 
\subsection{Density threshold and saturation}
\label{sec:density}
We apply a density threshold and a density saturation before the scattering scheme described in Sec. \ref{sec:scatter} by
\begin{equation}
\rho_{o'}(\boldsymbol r)=
\left\{ \begin{array}{ll}
  0, & \mbox{if $\rho_{o}(\boldsymbol r) < \rho_{\rm th}^{\rm low}$}; \\
  \rho_{\rm th}^{\rm high}, & \mbox{if $\rho_{o}(\boldsymbol r) > \rho_{\rm th}^{\rm high}$},
\end{array} \right.
\end{equation}
where $\rho_{o'}(\boldsymbol r)$ is the modified density, $\rho_{o}(\boldsymbol r)$ is the original ZA density, $\rho_{\rm th}^{\rm low}$ and $\rho_{\rm th}^{\rm high}$ are the density threshold and density saturation respectively.

One can adjust three-point statistics (e.g., bispectrum) by tuning the density threshold. 
An effective field theory description of the large-scale structure has been studied to constrain the dark matter three-point statistics (see \citealt{Angulo:2014tfa,Baldauf:2014qfa}).
Thereafter, \cite{Kitaura:2014mja} found that a density threshold is essential to accurately model the halo three-point statistics. 
The density saturation can be used to adjust the amplitude of the power spectrum. 
It can be considered as modifying the weights of the scatter for different density regions, i.e., the density regions above the density saturation will have the same weight during the scatter process.
While the scatter parameter for the PDF mapping can also be used to adjust the amplitude of the power spectrum, we find that the amplitude of the power spectrum is more sensitive to the density saturation. Therefore, we fix the scatter parameter to a constant and vary the density saturation only.

\subsection{Nonlinear effects correction}
\label{sec:nonlinear}
Nonlinear evolution of the dark matter density field and halo bias have partially been accounted for by the steps described above.  We correct for residual nonlinear effects by modifying the input power spectrum. 
In addition, our approximate method might introduce uncertainties of a few Mpc in the position of haloes, which will reduce the power at small scales and the signal of baryon acoustic oscillations (BAO).
The correction includes enhancing the BAO feature, and overall power spectrum at small scales.

The BAO is enhanced by
\begin{equation}
P_{\rm eBAO}(k)=(P_{\rm lin}(k)-P_{\rm nw}(k))\exp(k^2/k_*^2) + P_{\rm nw}(k),
\end{equation}
where $P_{\rm eBAO}(k)$ is the BAO enhanced power spectrum, $P_{\rm lin}(k)$ is the linear power spectrum, $P_{\rm nw}(k)$ is the smoothed no-wiggle power spectrum obtained by applying a cubic spline fit to $P_{\rm lin}(k)$, and $k_*$ is usually known as the damping factor (however, for the damping model, one should use $\exp(-k^2/k_*^2)$ instead).

We also enhance the power spectrum at small scales by
\begin{equation}
P_{\rm ePK}(k)=P_{\rm eBAO}(k)\cdot(1+Ak),
\end{equation}
where $A$ is a free parameter.

\subsection{Peculiar velocity}
\label{sec:velocity}
We model peculiar motions $v$ by adding to the linear coherent motion, which is proportional to the ZA displacement field, a dispersion term modeled by a random Gaussian distribution, i.e., 
\begin{equation}
v_i(\boldsymbol r)=B\psi_i(\boldsymbol r)+G(\lambda'),
\end{equation}
where $B$ is a constant corresponding to linear growth; $\psi$ is the displacement field, $i$ denotes the direction $x$, $y$, or $z$; and $G(\lambda)$ is a random number drawn from the Gaussian distribution with width $\lambda'$.

\subsection{Mass assignment}
\begin{figure}
\centering
\includegraphics[width=0.5\textwidth]{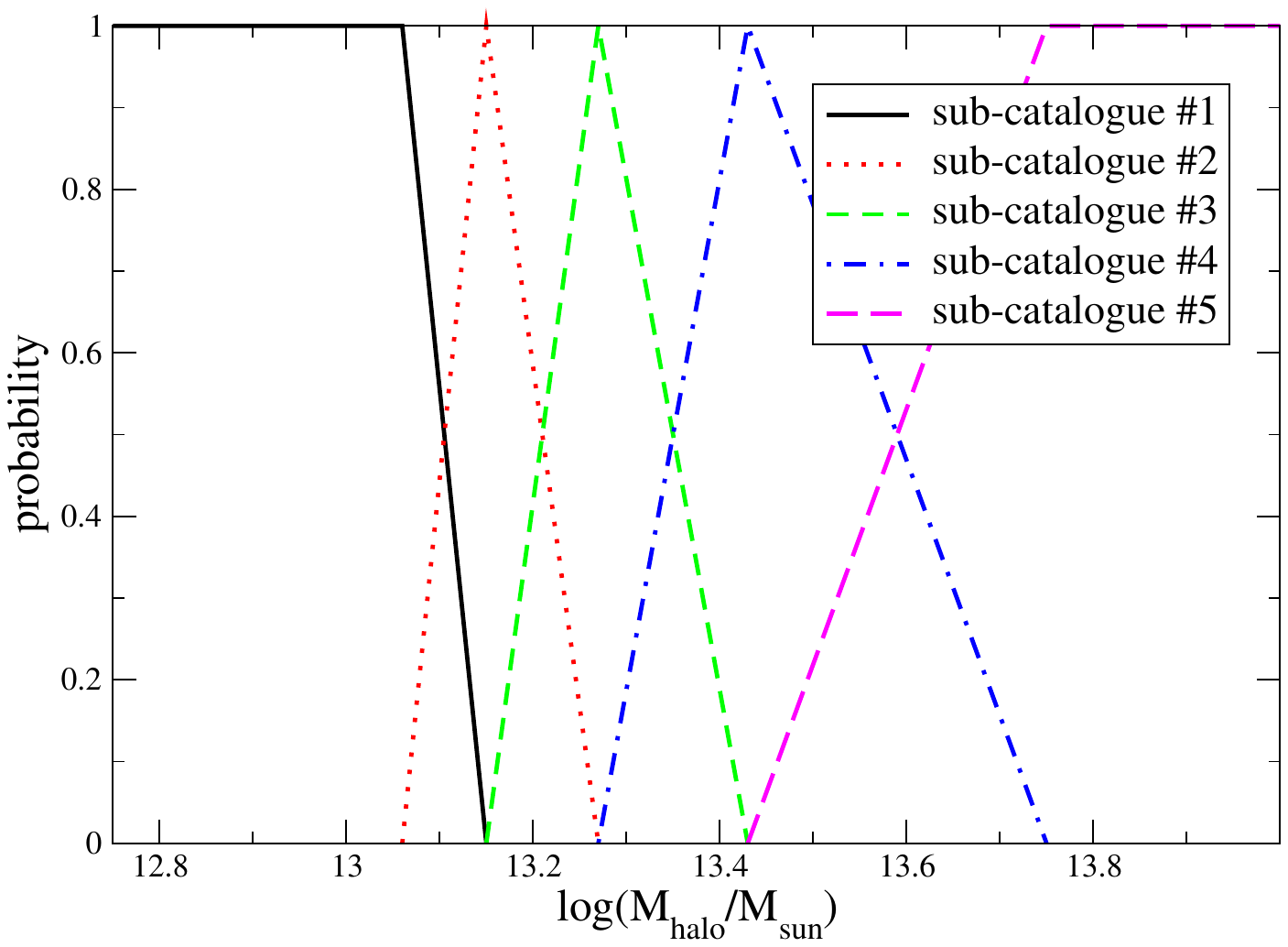}
\caption{The probability of each halo in the BigMD catalogue to be collected by one of the five sub-catalogues, represented by different colours. $M_{halo}$ is the halo mass in units of solar mass. These sub-catalogues are prepared to construct the EZmock including masses. The divisions are chosen, so that the sub-catalogues have a comparable number of haloes.}
\label{fig:masscic}
\end{figure}

We describe here how to assign masses to the distribution of tracers obtained in the previously described procedure with a simple additional step. First, we divide the BigMD halo catalogue into multiple catalogues according to the halo masses. Instead of taking the traditional mass bins (sharp mass cuts), we divide the sample in a statistical way. 
A halo will be place into one of the sub-catalogues with certain probability as shown in Fig. \ref{fig:masscic}. For example, a halo with $10^{13.6}$ solar mass has about half a chance to be collected by sub-catalogue \#4. If it is not collected by sub-catalogue \#4, it will be collected by sub-catalogue \#5. For each sub-catalogue, we construct an EZmock catalogue that will reproduce the clustering of each sub-catalogue. For each EZmock, we assign the halo mass with the value randomly selected from the sub-catalogue to which it corresponds. We demonstrate in the next section that the combined EZmock catalogue comprising all the sub-catalogues for different mass bins smoothly depends on arbitrary mass cuts. 

\section{Validation of the method} \label{sec:results}

In this section we validate our method comparing different statistics between EZmock and BigMD. The statistics include probability distribution function, power spectrum and two-point correlation function (monopole and quadrupole), bispectrum, and mass function. 
We develop a best-fit finder algorithm to find the best-fit parameters, $\rho^{low}_{th}=2.4$ (per cell),  $\rho^{high}_{th}=3.5$ (per cell), $k_*=0.09 \hompc$, $A=0.07 \mpcoh$, and $B=75$ {\rm km} $s^{-1}\hompc$. The performance is shown in following subsections.

\subsection{Probability distribution function}
\label{sec:pdf_result}

\begin{figure}
\centering
\includegraphics[width=0.5\textwidth]{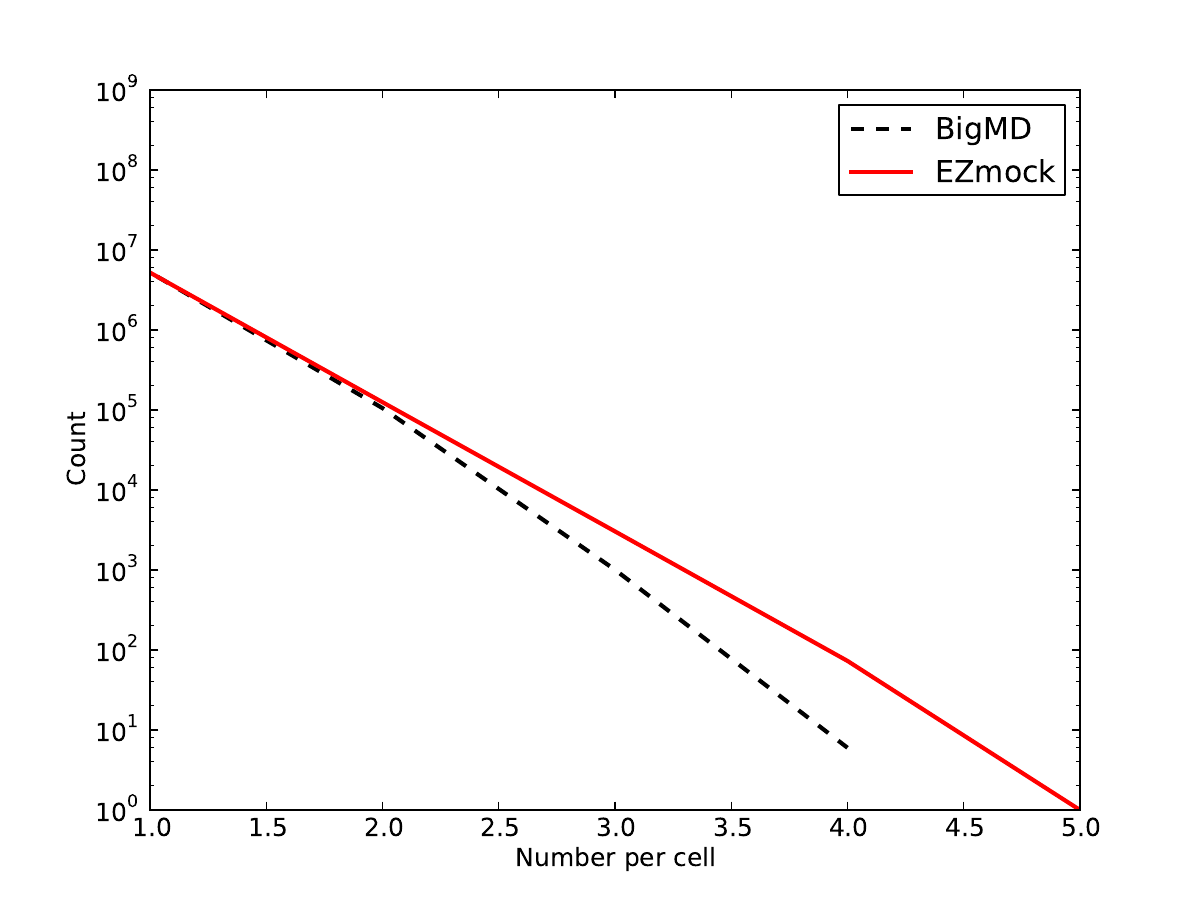}
\caption{Halo probability distribution functions for EZmock (solid red line) and BigMD (dashed black line). The PDFs are computed using nearest-grid-point (NGP) with a grid of $960^3$.}
\label{fig:pdf}
\end{figure}

The PDF mapping ensures that the integral of the PDF, i.e., the number density, is fitted by construction.
It also encodes information on the higher-order correlation functions (see, e.g., \citealt{Kitaura:2014mja}). Therefore, an accurate mock should also reproduce the shape of the PDF. We note, however, that it is difficult to perfectly match the tail of the PDF towards large numbers of haloes due to the small number statistics in such bins. Fig. \ref{fig:pdf} shows the PDFs of EZmock and BigMD. The PDFs are computed using nearest-grid-point (NGP) with a grid of $960^3$. 

Since the haloes are populated using CIC (see Sec. \ref{sec:PDF}) to generate a continuous halo number density field,
it is not guaranteed that the reference PDF is restored in the PDF mapping. Nevertheless, the PDF is sufficiently matched to ensure that the higher-order statistics is accurately reproduced, as we demonstrate below.

\subsection{Power spectrum and two-point correlation function}
\label{sec:pk_2pcf}

\begin{figure*}
\begin{center}
 \subfigure{\label{fig:pk_r}\includegraphics[width=0.49 \textwidth]{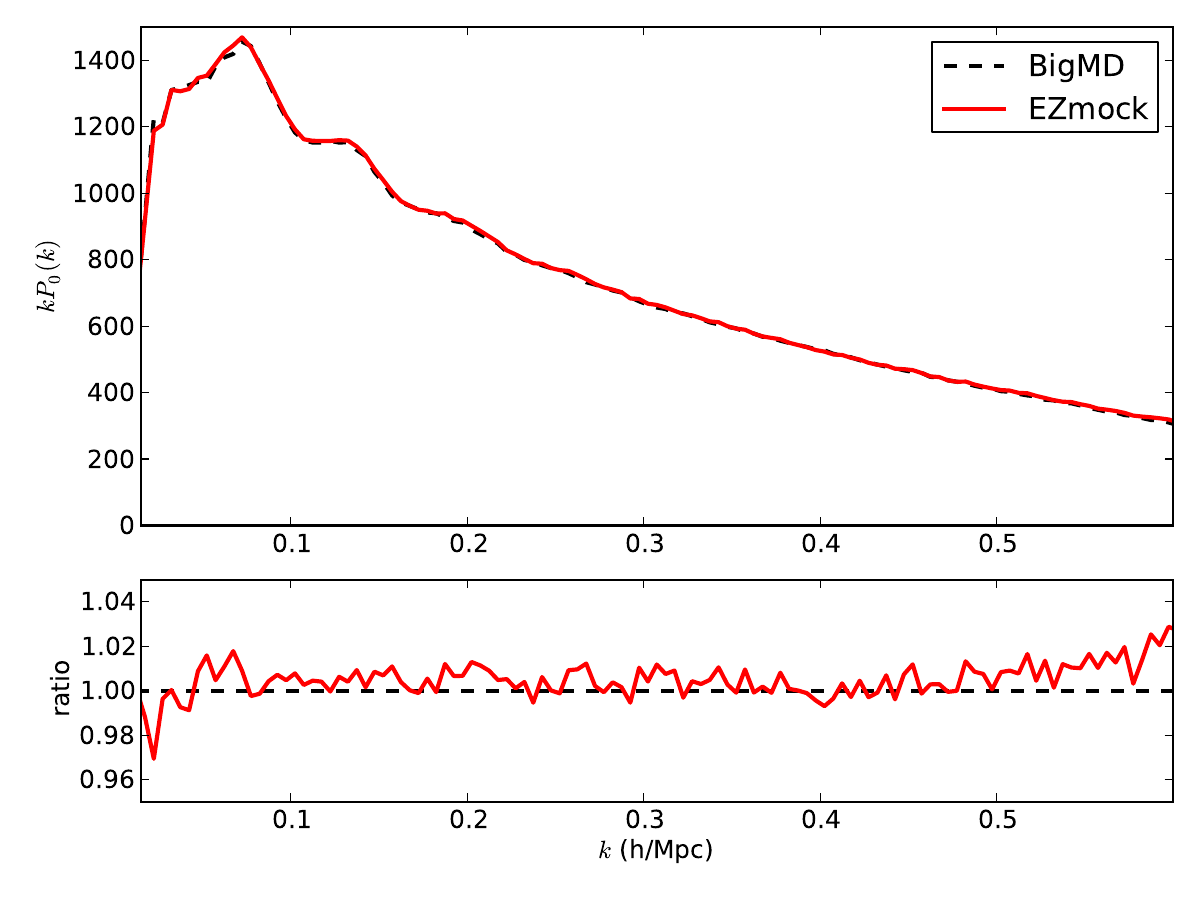}}
 \subfigure{\label{fig:pk_z}\includegraphics[width=0.49 \textwidth]{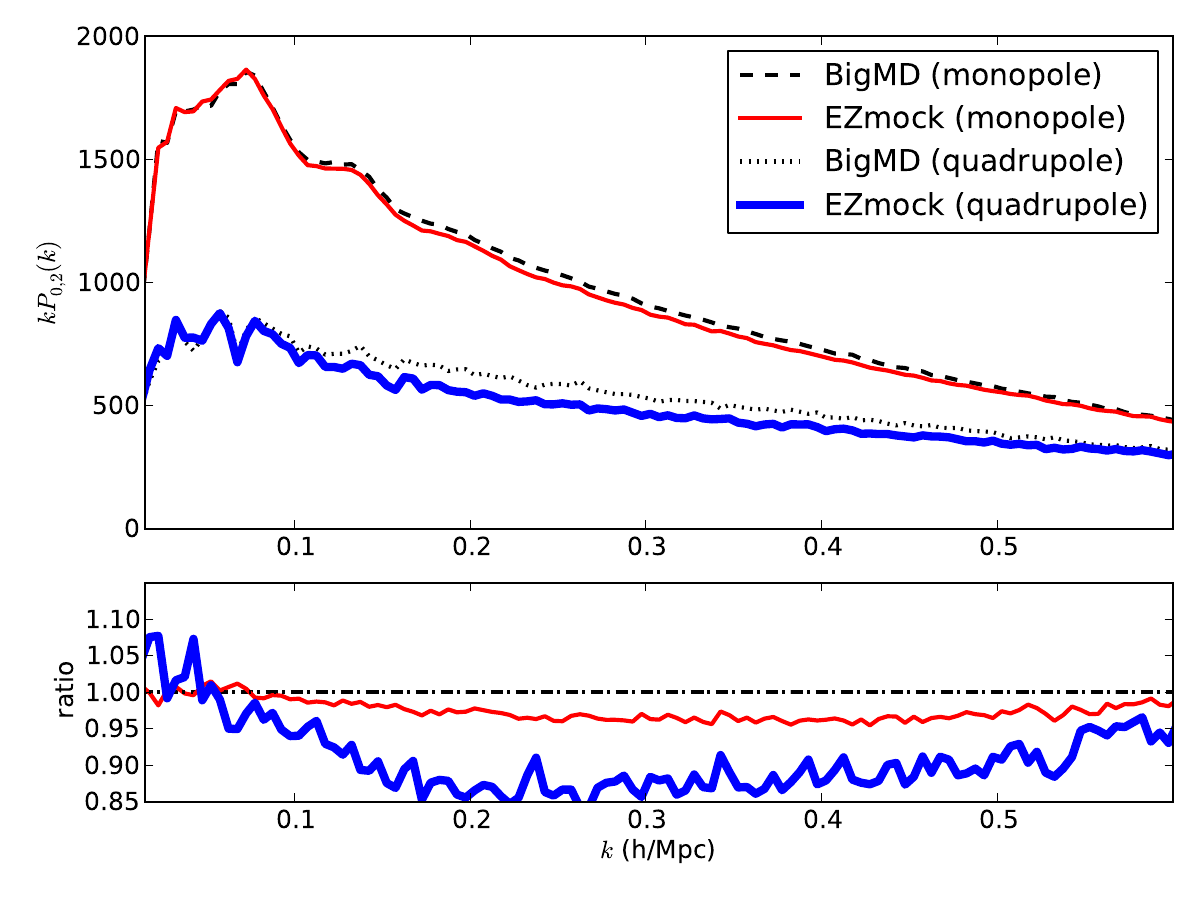}}
\end{center}
\caption{Left panel: monopole of the power spectrum in real space. Right panel: monopole and quadrupole of the power spectrum in redshift-space.}
\label{fig:pk}
\end{figure*}

One of the main goals of  mock galaxy catalogues is to accurately reproduce the theoretically expected two-point statistics (power spectrum in Fourier-space), including cosmic variance, i.e., the uncertainties due to the particular realisation of seed perturbations.  When using Gaussian or log-normal realisations, one is able to impose an arbitrary power spectrum.

However, modelling gravitational evolution with perturbation theory introduces a bias, due to the approximate modelling of the nonlinear regime with respect to full gravity solutions. This is seen in the power spectrum as a damping of the small scale structures towards high $k$s. Therefore, some works suggested to use a transfer function connecting the full with the perturbative solution (see \citealt{Tassev:2012cq,Tassev:2011ac}). Yet, this transfer function introduces ringing effects in the fields coming from Fourier-space deconvolution effects and hence, it cannot be applied to produce mock catalogues.
It was therefore suggested in \cite{Kitaura:2013cwa} to include the perturbation theory induced damping effect in the bias. We follow this approach here obtaining remarkable results.. 

Fig. \ref{fig:pk} (left panel) shows the comparison of the monopole of the power spectrum between the EZmock and BigMD in real-space. They agree within $1\%$ accuracy in the scale range up to $k=0.55$ $\hompc$. Since we generate the EZmock halo catalogues with a finite grid ($960^3$), it is not meaningful to compare the power spectrum at the scales which are close to the Nyquist frequency, i.e., at $k>0.6$ $\hompc$. It would be interesting to explore the validity of our methodology at smaller scales, however, such a study is out of the scope of this paper.

Fig. \ref{fig:pk} (right panel) shows the monopole and quadrupole of the power spectrum in redshift-space. Our simple model (linear + Gaussian, see Sec. \ref{sec:velocity}) to assign velocities to haloes has a limited performance, yielding a worse fit than in real-space. However, while there is still some room for improvement, it should be already good enough for most of practical applications, as the Kaiser factor is well recovered.

\begin{figure*}
\begin{center}
 \subfigure{\label{fig:cf_r}\includegraphics[width=0.49 \textwidth]{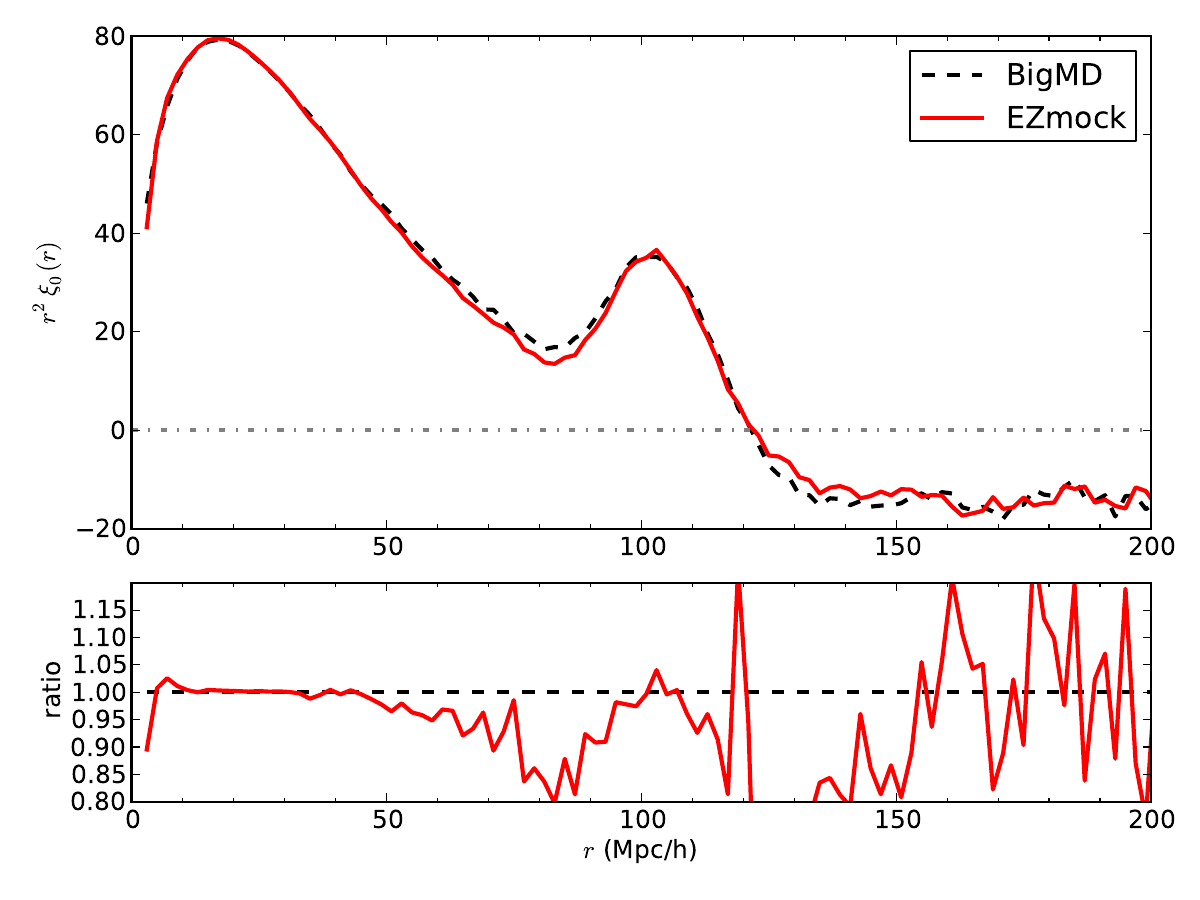}}
 \subfigure{\label{fig:cf_z}\includegraphics[width=0.49 \textwidth]{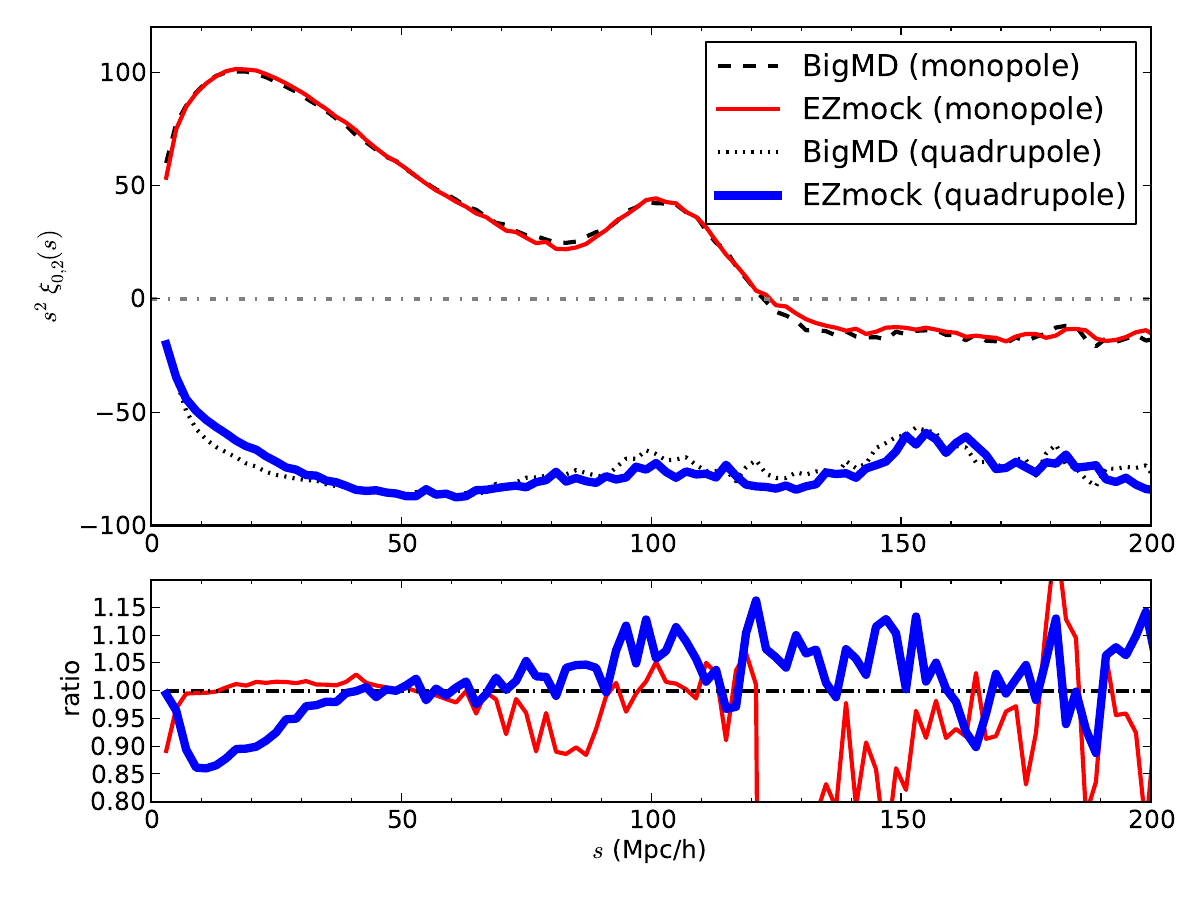}}
\end{center}
\caption{Left panel: monopole of the two-point correlation function in real space. Right panel: monopole and quadrupole of the two-point correlation function in redshift-space.}
\label{fig:cf}
\end{figure*}

This can be seen in Fig. \ref{fig:cf}, where the monopole and quadrupole of the two-point correlation function in real and redshift-space are shown.  The correlation function is restored down to $10$ $\mpcoh$ within $1\%$.

\subsection{Bispectrum}
\label{sec:bik}

\begin{figure*}
\begin{center}
 \subfigure{\label{fig:bik0d05}\includegraphics[width=0.49 \textwidth]{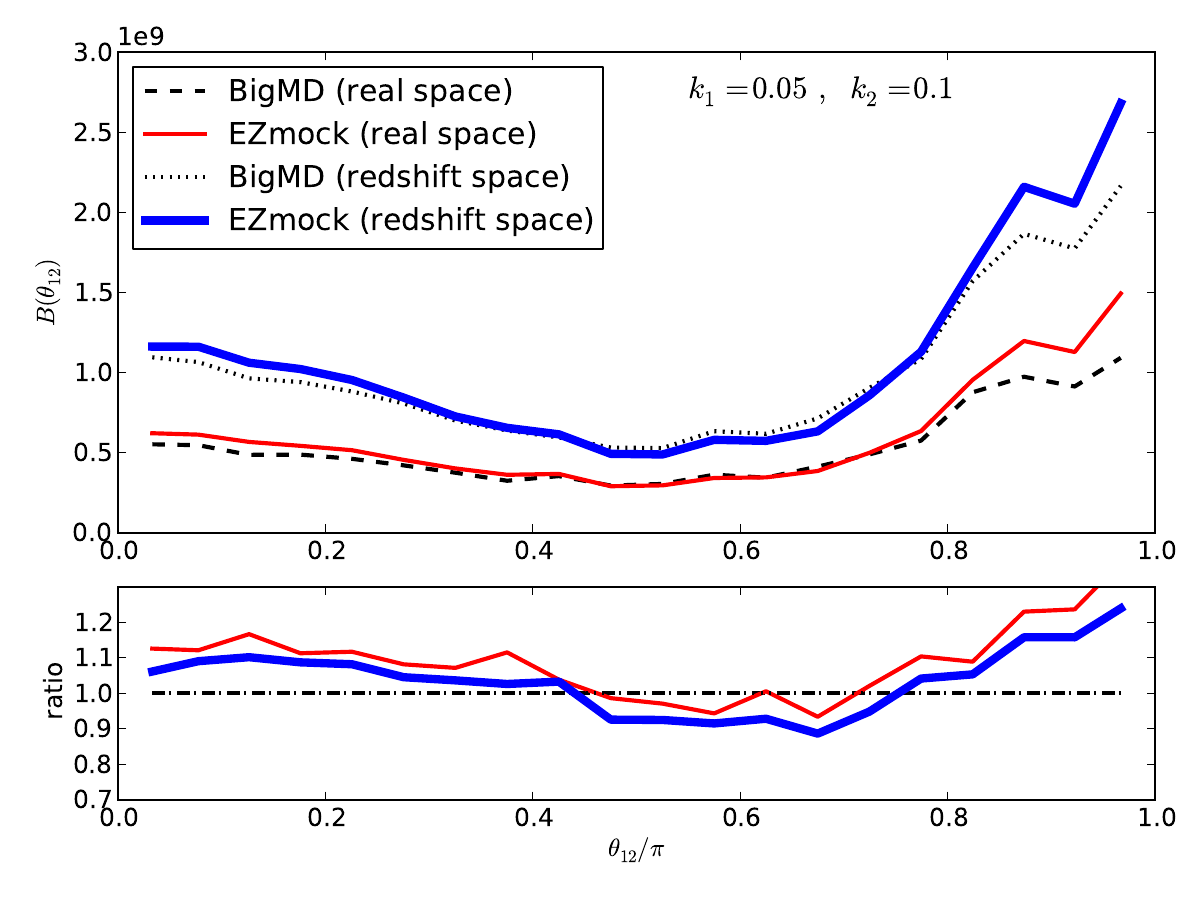}}
 \subfigure{\label{fig:bik0d1}\includegraphics[width=0.49 \textwidth]{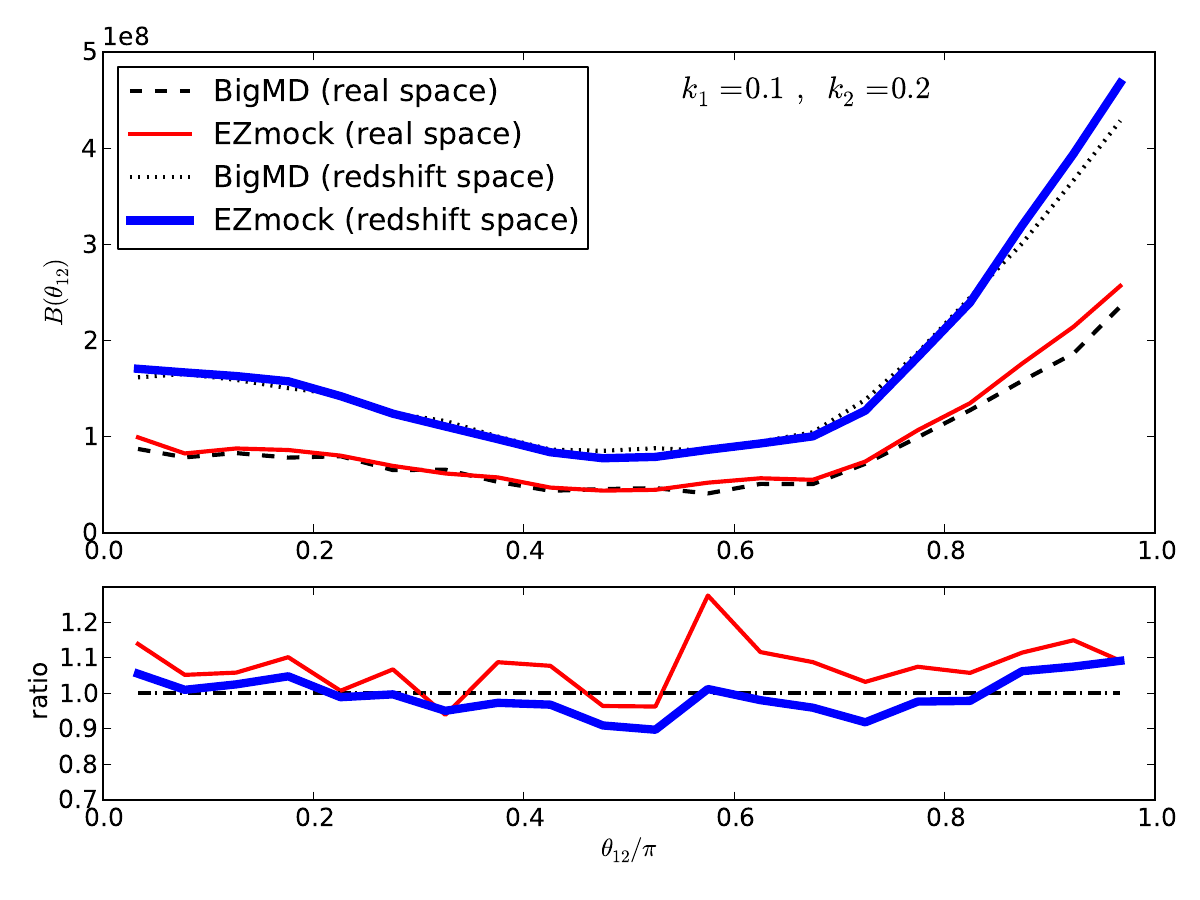}}
 \subfigure{\label{fig:bik0d15}\includegraphics[width=0.49 \textwidth]{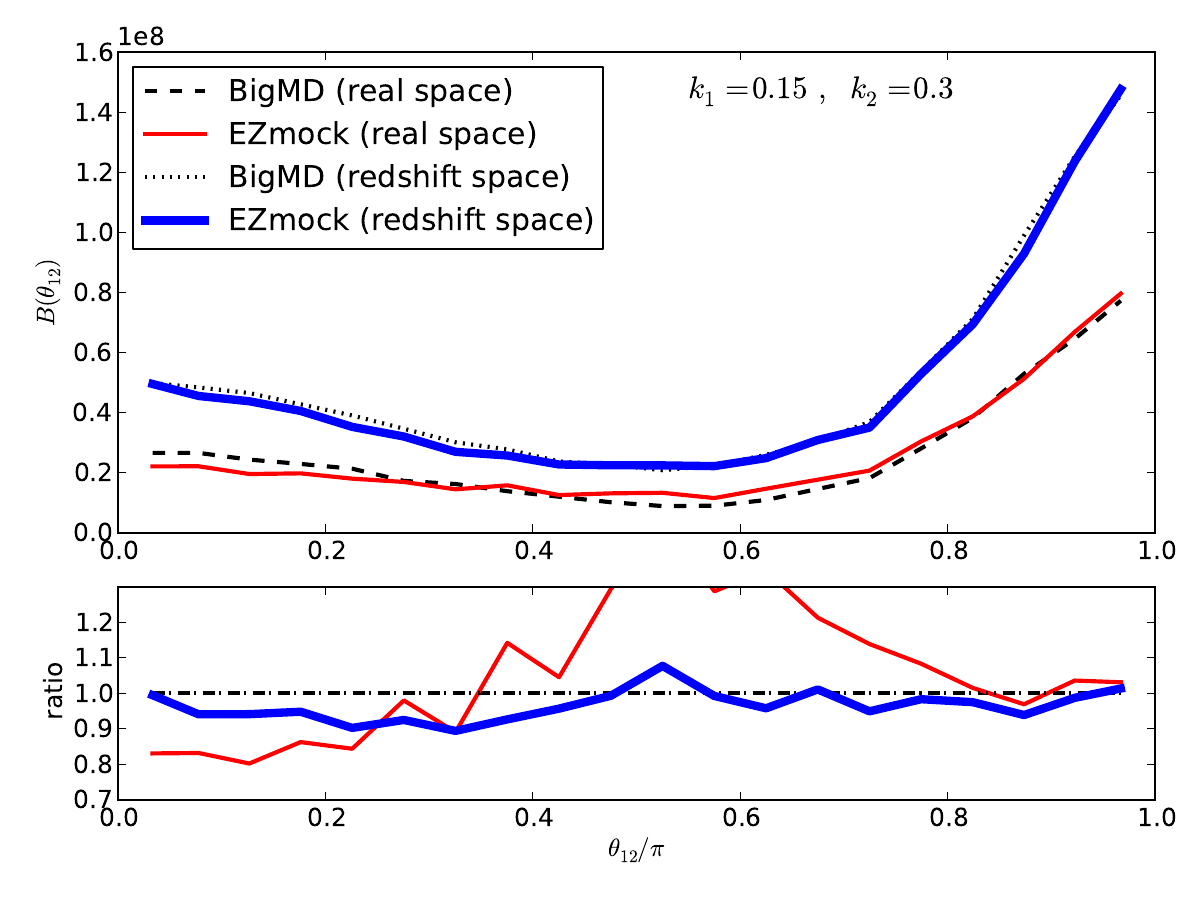}}
 \subfigure{\label{fig:bik0d2}\includegraphics[width=0.49 \textwidth]{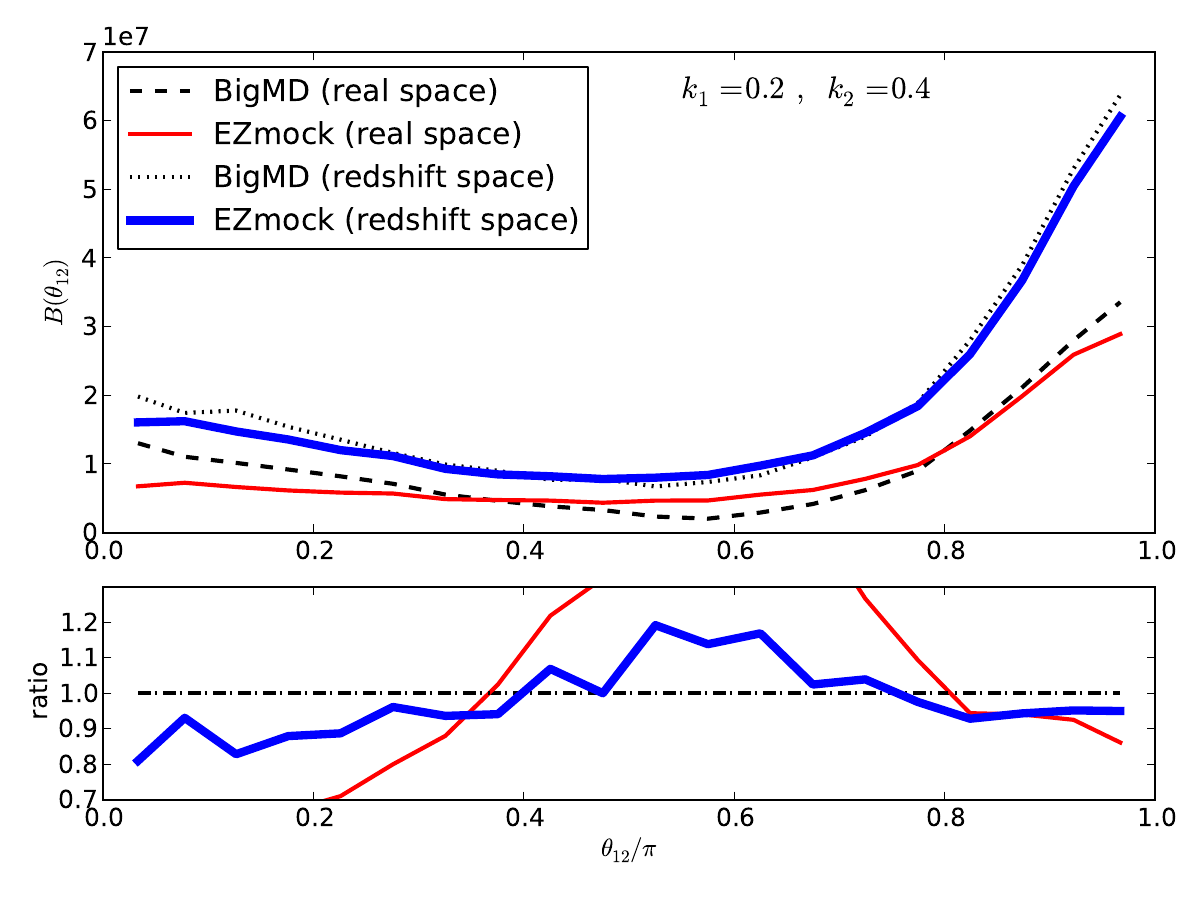}}
\end{center}
\caption{Bispectrum in real and redshift-space. The configurations including $k_2=2k_1=0.1$, $k_2=2k_1=0.2$, $k_2=2k_1=0.3$, and $k_2=2k_1=0.4$ $\hompc$ are denoted in the panels.}
\label{fig:bik}
\end{figure*}

We find that, in general terms, the three-point statistics is well modelled by the effective Zel'dovich approximation on large scales.
Fig. \ref{fig:bik} shows the bispectrum of different configurations, $k_2=2k_1=0.1$, $k_2=2k_1=0.2$, $k_2=2k_1=0.3$, and $k_2=2k_1=0.4$ $\hompc$, in real and redshift-space. 
The bispectrum is fit mostly within $20\%$ accuracy for the configurations, $k_2=2k_1=0.1$ and $k_2=2k_1=0.2$ $\hompc$, in real-space, and for all the configurations tested, i.e., $k_2=2k_1=0.1$, $k_2=2k_1=0.2$, $k_2=2k_1=0.3$, and $k_2=2k_1=0.4$ $\hompc$, in redshift-space.
The accuracy in the three point statistics could be improved using a more accurate structure formation model, such as Augmented Lagrangian Perturbation theory (ALPT, \citealt{Kitaura:2012tj}). This would have, of course, an impact on the computational efficiency of our method. We will leave such an improvement for a later work.

\subsection{Mass vs. bias}
\label{sec:mass}

\begin{figure*}
\begin{center}
\subfigure{\label{fig:massbin}\includegraphics[width=0.49\textwidth]{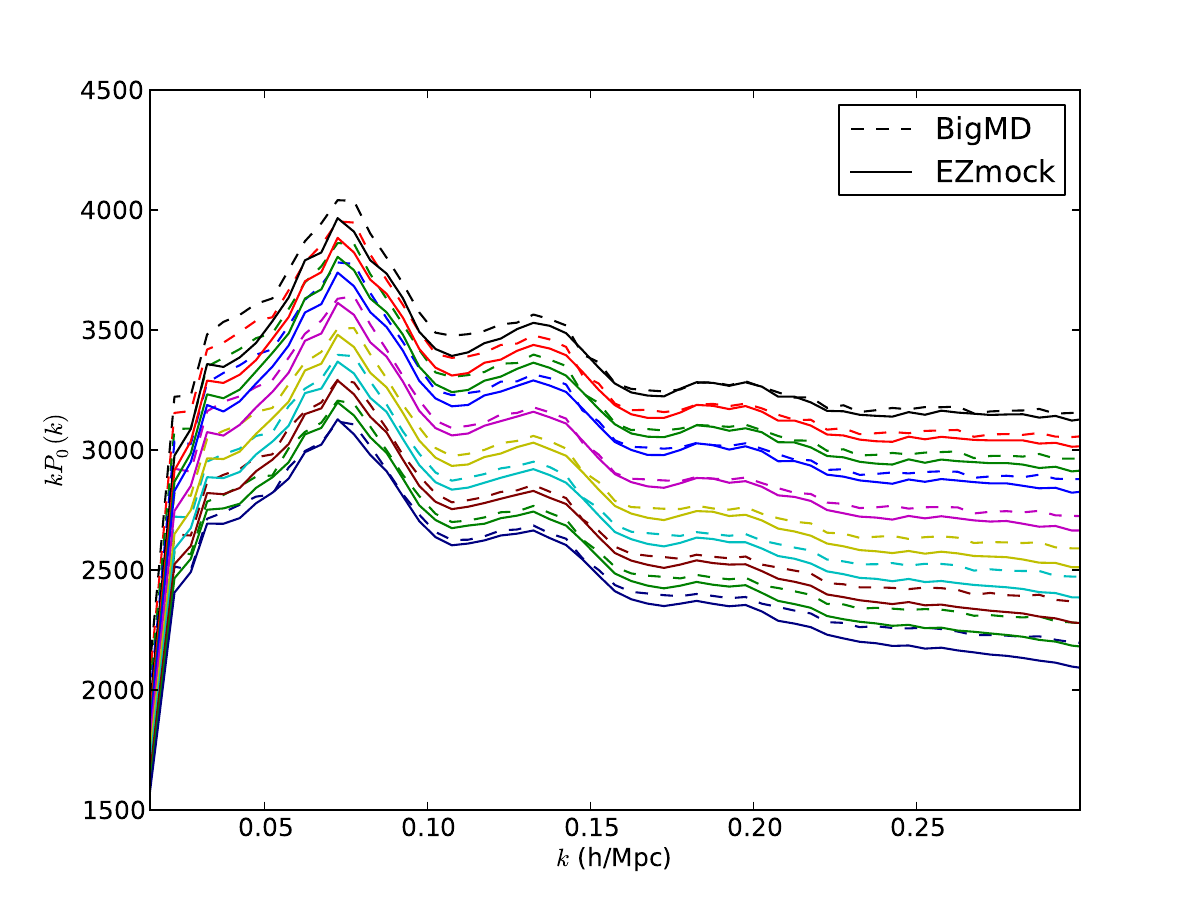}}
\subfigure{\label{fig:massfunc}\includegraphics[width=0.49\textwidth]{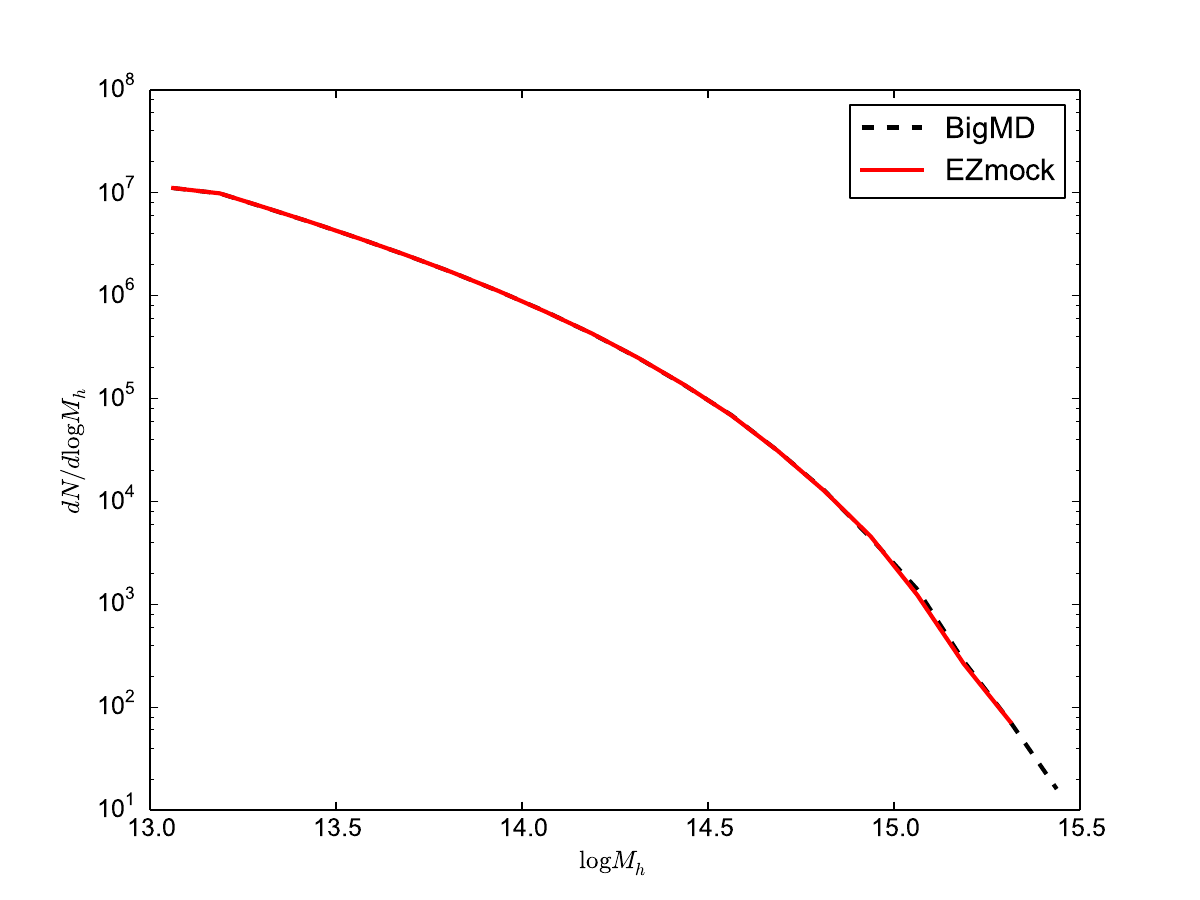}}
\end{center}
\caption{Left panel: power spectra in real space with different number densities, $\{2, 2.1, 2.2, 2.3. 2.5, 2.7, 2.9, 3.1, 3.3, 3.5\}\times10^{-4}$ $h^3\,{\rm Mpc}^{-3}$. Each subsample of a given number density is selected using a mass threshold, i.e., all the haloes with masses larger than a certain mass threshold  with the right number density are selected. Right panel: mass functions for the EZmock and BigMD catalogues. They agree by construction.
}
\label{fig:mass}
\end{figure*}

Fig. \ref{fig:mass} (left panel) shows the power spectrum in real-space for 10 different number densities, demonstrating that EZmock smoothly mimics the dependency between mass and bias using our methodology with 5 mass bins (see \ref{sec:mass}). One can still achieve a greater accuracy by increasing the number of mass bins. Fig. 6 (right panel) demonstrates that the mass function has been restored by construction.

\section{Conclusion and discussion}
\label{sec:conclusion}

In this work, we have developed a new methodology to construct the effective Zel'dovich approximation mock catalogues (EZmocks), which provide an efficient way to generate massive mock catalogues with accurate one-, two-, and three-point clustering statistics. We have shown the good performance of EZmock by comparing the measurements in real and redshift-space with that from a large $N$-body simulation, BigMD. Constructing one mock, using a grid of $960^3$ takes only a few minutes while using a machine with multicore processors (e.g., it takes 5 minutes while using 16 cores machine in our case). We have also presented an example to demonstrate the flexibility of assigning masses to EZmock, which gives a reasonable mass vs. bias relation.

We propose in this work to push the Zel'dovich approximation to its limits by extending it with effective models, which can compensate and alleviate the missing physical contributions. We demonstrate that such an approach is very promising to achieve high precisions, only reached with more complex approaches at higher computational costs.
Our effective model including 1) PDF mapping with scattering, 2) density threshold and saturation, and 3) enhancing BAO and power spectrum at small scales, effectively absorbs the effects/corrections due to the nonlinear growth of the density field and halo bias (including deterministic and stochastic biasing), so that EZmock can precisely reproduce the clustering statistics. While we have used the simplest model with the minimum number of parameters (e.g., step function) to capture the major characteristics of the effects/corrections, we expect that the performance of EZmock can be improved by using more precise models (e.g., using some smooth function instead of a step function). 

Our method should be regarded as complementary to the efforts in producing catalogues based on full gravity solutions. High quality $N$-body simulations will remain indispensable to produce accurate reference catalogues, which can be massively reproduced with methods like the one presented in this paper.
In addition, EZmock, at this stage, aims at providing a precise and practical tool for estimating the covariance matrices. It can be also conveniently used for future survey forecasts. EZmock provides a significant improvement with respect to log-normal mocks at a comparable computational cost.
 
In summary, provided the efficient production and accurate clustering statistics, EZmock can be one of the most practical methods to generate a large number of mock catalogues for analysing large-scale structure galaxy surveys.

\section{Acknowledgement}
We thanks the reviewer, David Weinberg, for useful comments. 
CC and FP were supported by the Spanish MICINN’s Consolider-Ingenio 2010 Programme under grant MultiDark CSD2009-00064 and AYA2010-21231-C02-01 grant, the Comunidad de Madrid under grant HEPHACOS S2009/ESP-1473, and Spanish MINECO’s “Centro de Excelencia Severo Ochoa” Programme under grant SEV-2012-0249.  CZ acknowledges support from Charling Tao and her grant from Tsinghua University, and 973 program No. 2013CB834906. CZ also thanks the support from MultiDark summer student program to visit the Instituto de F\'{\i}sica Te\'orica, (UAM/CSIC), Spain.
GY acknowledge support from the Spanish MINECO under research grants  AYA2012-31101, FPA2012-34694, AYA2010-21231, Consolider Ingenio SyeC CSD2007-0050 and  from Comunidad de Madrid under  ASTROMADRID  project (S2009/ESP-1496).

The MultiDark Database used in this paper and the web application providing online access to it were constructed as part of the activities of the German Astrophysical Virtual Observatory as result of a collaboration between the Leibniz-Institute for Astrophysics Potsdam (AIP) and the Spanish MultiDark Consolider Project CSD2009-00064.  The  BigMD simulation suite have been performed in the Supermuc supercomputer at LRZ using time granted by PRACE.

We acknowledge PRACE for awarding us access to resource Curie supercomputer based in France.

%\begin{thebibliography}{}

%\end{thebibliography}

\label{lastpage}

{\small
\bibliography{ezmock}
}

\end{document}